\documentclass[11pt]{amsart}

\title[Superposition and time delay are incompatible]
{Superposition and time delay are incompatible between them}
\author{Eduardo S. Zeron}
\address{Departamento de Matem\'aticas, CINVESTAV del IPN, 
Apartado Postal 14-740, Ciudad de M\'exico, 07000, M\'exico.}
\email{eszeron@math.cinvestav.edu.mx}

\date{\today}
\thanks{Research supported by Cinvestav del IPN and Conacyt in M\'exico.}

\begin{document}
\begin{abstract}
We analyse the properties of wave functions $\psi(x,t;y,s)$ that 
depend at once on two time-position vectors $(x,t)$ and $(y,s)$ with 
different time entries $t\neq{s}$. In particular, we prove that the
existence of such wave functions is indeed incompatible with the 
superposition property and the technique of separation of variables.
\end{abstract}
\maketitle

\section{Time delayed wave functions}

Quantum mechanics is one of the most successful theories of the twenty 
century; its predictions have been verified with such a remarkable 
precision, that one is automatically compelled to consider whether the 
Schr\"odinger equation can be modified to include different phenomena, 
like time delays. Many works can be found in the literature on this 
subject; see for example Muga et al's books \cite{Muga-2008,Muga-2009}. 
Moreover, in chapter xviii of B\"ohm's book \cite{Boehm-1986} it is 
analysed the relation between time delays and quantum scattering; while 
in chapters 18 and 19 of Razavy's book \cite{Razavy-2003} it is analysed 
the connection between time delays and quantum tunneling. In any case, 
we can safely assert that time delays and quantum mechanics seem to have 
a strange and intriguing relationship, that maybe goes back to 1978, 
when Wheeler proposed his delayed choice \textit{gedanken} experiment 
\cite{Wheeler-1983,Lloyd-2012}. This relation became even stronger in 
2000 and later, when the delayed choice quantum eraser was 
experimentally verified \cite{Kim-al-2000,Jacques-al-2007,
Scarcelli-al-2007,Peruzzo-al-2012,Kaiser-al-2012}.

Nevertheless, the connections between time delays and quantum mechanics 
mentioned above seem to be subjective, so that we wonder whether they 
can be formalised; i.e., we wonder whether there could exists a wave 
function $\psi^*(x,t;y,s)$ that depends at once on two time-position 
vectors $(x,t)$ and $(y,s)$ with different time entries $t\neq{s}$. We 
are in particular interested in the simple case, when the difference 
$t-s$ is equal to a fixed constant $\tau>0$, so that the square of the 
absolute value $|\psi^*(x,t;y,s)|^2$ is the probability density function 
that the system \textbf{is} in the position $x$ at time $t$ and it 
\textbf{was} in the position $y$ at time $s=t-\tau$.

Now then, if such a wave function $\psi^*(x,t;y,s)$ exists, the state 
of the system would be described by two objects: a classical wave 
function $\psi(x,t)$ and the new one $\psi^*$. In particular, the 
term $i\hbar\frac\partial{\partial{t}}\psi(x,t)$ in the Schr\"odinger 
equation should be equal to a Hamiltonian that linearly depends on 
one or all of the following functions: $\psi(x,t)$, $\psi(y,s)$, 
and $\psi^*(x,t;y,s)$.

The main objective of this work is to prove that the existence of such a 
wave function $\psi^*(x,t;y,s)$ is incompatible with the superposition 
property and the technique of separation of variables; and this 
incompatibility does not come from the structure of the corresponding 
Schr\"odinger equation, but from the fact that $|\psi^*(x,t;y,s)|^2$ is 
the probability density function that the system is in the state $x$ at 
time $t$ and it was (or will be) in the state $y$ at time $s$.
 
Since $|\psi(x,t)|^2$ is the probability density function that the 
system is in the state $x$ at time $t$, its integral with respect 
to $x$ must be equal to one. In the same way, the integral of 
$|\psi^*(x,t;y,s)|^2$ with respect to the variable $y$ must be equal 
to the probability that the system is at state $x$ at time $t$; i.e., 
we have for all times $t$ and $s$ that
\begin{eqnarray}\label{eqn1} 
\int\big|\psi(x,t)\big|^2dx&\equiv&1\quad\hbox{and}\\ 
\label{eqn2}\int\big|\psi^*(x,t;y,s)\big|^2dy 
&=&\big|\psi(x,t)\big|^2. 
\end{eqnarray}

We prove in the following two chapters that the previous pair of 
identities is indeed incompatible with the superposition property 
and the technique of separation of variables.

\section{Separation of variables and superposition}

The fact that the system could be described by a pair of wave functions 
$\psi^*$ and $\psi$ drives us to think that there should be a relation 
between them; i.e. there should exist an operator that transforms the 
function $\psi^*(x,t;y,s)$ into $\psi(x,t)$ by \textit{forgetting} the 
dependence on the time-position vector $(y,s)$. One of the simplest 
forms to visualise this relation is to use the technique of separation 
of variables, which asserts that one can find wave functions (solutions 
to the corresponding Schr\"odinger equation) of the form
\begin{equation}\label{eqn3}
\psi^*_k(x,t;y,s)=\psi_k(x,t)\xi_k(y,s) 
\end{equation}
for some complex function $\xi_k(y,s)$. Since $\psi^*_k=\psi_k\xi_k$ 
must satisfy the identity (\ref{eqn2}), one obviously have that
\begin{equation}\label{eqn4}
\int\big|\xi_k(y,s)\big|^2dy\equiv1\quad\hbox{for every time}\;s.
\end{equation}

Nevertheless, not every wave function $\psi^*(x,t;y,s)$ can be 
decomposed into a product like in (\ref{eqn3}). Consider for example 
two wave functions $\psi^*_k=\psi_k\xi_k$ that satisfy (\ref{eqn3}) for 
$k=1,2$. Given a pair of non-zero constants $\beta_1$ and $\beta_2$, it 
is easy to verify that the superposition 
$\beta_1\psi^*_1{+}\beta_2\psi^*_2$ has a product decomposition as in 
(\ref{eqn3}) only when there is a constant $\gamma\neq0$ such that 
$\psi_2\equiv\gamma\psi_1$ or $\xi_2\equiv\gamma\xi_1$.

Similar restrictions on $\psi_k$ and $\xi_k$ can be deduced, when one 
requests that the superpositions $\beta_1\psi^*_1{+}\beta_2\psi^*_2$ 
and $\beta_1\psi_1{+}\beta_2\psi_2$ satisfy the identity (\ref{eqn2}). 
We must point out that $\beta_1\psi^*_1{+}\beta_2\psi^*_2$ does not need 
to have a product decomposition, in order to obtain the new restrictions. 
Thus, let $\psi^*_k=\psi_k\xi_k$ be two wave functions that satisfy the 
corresponding Schr\"odinger equation and the identities (\ref{eqn2}) 
and (\ref{eqn3}) for $k=1,2$. The linear combination 
$\beta_1\psi^*_1{+}\beta_2\psi^*_2$ obviously satisfy the same 
Schr\"odinger equation.  Moreover, the following identity easily holds, 
$$\big|\beta_1\psi^*_1+\beta_2\psi^*_2\big|^2 
=|\beta_1\psi^*_1|^2+2\Re\Big(\beta_1\psi^*_1 
\overline{\beta_2\psi^*_2}\Big)+|\beta_2\psi^*_2|^2;$$ 
and a similar result is obtained by writing $\psi_k$ 
instead of $\psi^*_k$. Whence, the linear combinations 
$\beta_1\psi^*_1{+}\beta_2\psi^*_2$ and $\beta_1\psi_1{+}\beta_2\psi_2$ 
satisfy the identity (\ref{eqn2}) if and only if: 
$$\Re\Big(\beta_1\overline{\beta_2}\int\!\psi^*_1(x,t;y,s) 
\overline{\psi^*_2(x,t;y,s)}dy\Big)=\Re\Big(\beta_1 
\overline{\beta_2}\psi_1(x,t)\overline{\psi_2(x,t)}\Big).$$

One can easily rewrite this equality by recalling equation (\ref{eqn3}) 
and introducing the inner product $\langle\xi_1,\xi_2\rangle_s$ that 
depends on $s$,
\begin{eqnarray}\label{eqn5}
\Re\Big(\beta_1\overline{\beta_2}\psi_1(x,t)\overline{\psi_2(x,t)}
\big(\langle\xi_1,\xi_2\rangle_s{-}1\big)\Big)=0,&&\\
\label{eqn6}\hbox{where}\qquad{s}\mapsto\langle\xi_1,
\xi_2\rangle_s:=\int\!\xi_1(y,s)\overline{\xi_2(y,s)}dy.&&
\end{eqnarray}

We can so conclude that there exists a pair of non-zero constants 
$\beta_1$ and $\beta_2$ for which the identities (\ref{eqn2}) and 
(\ref{eqn5}) hold if and only if one of the following conditions is 
satisfied:
\begin{enumerate} \setcounter{enumi}{6}
\item\label{eqn7} The inner product 
$\langle\xi_1,\xi_2\rangle_s\equiv1$ for every time $s$.
\item\label{eqn8} There are two real constants 
$\theta_1$ and $\theta_2$ such that the functions 
$\hbox{e}^{-i\theta_1}\psi_1(x,t)\overline{\psi_2(x,t)}\in\mathbb{R}$ and 
$\hbox{e}^{-i\theta_2}\big(\langle\xi_1,\xi_2\rangle_s{-}1\big)\in\mathbb{R}$
take only real values for all positions $x$ and times $t$ and $s$. 
\end{enumerate}

Notice that the latter condition (\ref{eqn8}) automatically implies 
that one can find constants $\beta_1$ and $\beta_2$ such that 
$\hbox{e}^{i\theta_1+i\theta_2}\beta_1\overline{\beta_2}$ is purely 
imaginary, and so the identities (\ref{eqn5}) and (\ref{eqn2}) hold. 

\section{Conclusions}

This result has been proved in the previous chapter: If there exists a 
pair of wave functions $\psi^*_k=\psi_k\xi_k$ that satisfy the identities 
(\ref{eqn2}) and (\ref{eqn3}) for $k=1,2$, then the superpositions 
$\beta_1\psi^*_1{+}\beta_2\psi^*_2$ and $\beta_1\psi_1{+}\beta_2\psi_2$ 
also satisfy the identity (\ref{eqn2}) if and only if the condition 
(\ref{eqn8}) holds or the inner product $\langle\xi_1,\xi_2\rangle_s\equiv1$ 
for every time $s$. Both conditions obviously impose strong restrictions
on the functions $\psi_k$ and $\xi_k$, so that we can safely conclude 
that the existence of such wave functions $\psi^*_k=\psi_k\xi_k$ is 
incompatible with the superposition property and the technique of 
separation of variables.

\addcontentsline{toc}{section}{References}

\bibliographystyle{siam}
\bibliography{noise1}

\end{document}